\documentclass[pra,twocolumn,preprintnumbers,amsmath,amssymb,nofootinbib,floatfix]{revtex4}

\usepackage{graphicx, bm, tikz}
\usepackage[breaklinks]{hyperref}

\makeatletter
\def\graphicscale{\twocolumn@sw{0.3}{0.4}}
\def\graphicthreescale{\twocolumn@sw{0.3}{0.4}}

\begin{document}

\title{Scaling behavior of the stationary states arising from dissipation\\
 at continuous quantum transitions}

\author{Davide Rossini}
\affiliation{Dipartimento di Fisica dell'Universit\`a di Pisa
        and INFN, Largo Pontecorvo 3, I-56127 Pisa, Italy}

\author{Ettore Vicari} 
\affiliation{Dipartimento di Fisica dell'Universit\`a di Pisa
        and INFN, Largo Pontecorvo 3, I-56127 Pisa, Italy}

\date{\today}

\begin{abstract}
  We study the critical behavior of the nonequilibrium dynamics and of
  the steady states emerging from the competition between coherent and
  dissipative dynamics close to quantum phase transitions.  The latter
  is induced by the coupling of the system with a Markovian bath, such
  that the evolution of the system's density matrix can be effectively
  described by a Lindblad master equation.  We devise general scaling
  behaviors for the out-of-equilibrium evolution and the stationary
  states emerging in the large-time limit for generic initial
  conditions, in terms of the parameters of the Hamiltonian providing
  the coherent driving and those associated with the dissipative
  interactions with the environment.  Our framework is supported by
  numerical results for the dynamics of a one-dimensional lattice
  fermion gas undergoing a quantum Ising transition, in the presence
  of dissipative mechanisms which include local pumping and decay of
  particles.
\end{abstract}

\maketitle

% ========================= BODY =========================

\section{Introduction}
\label{intro}

One of the major challenges of current experimental and theoretical
investigations in the field of quantum statistical mechanics and
condensed matter physics is the understanding of the
out-of-equilibrium dynamics of open many-body systems, arising from
coherent Hamiltonian drivings and dissipative mechanisms.  The recent
technological breakthroughs in the manipulation of atomic and quantum
optical systems are paving the way to a careful study of the interplay
between the coherent quantum dynamics and the dissipative effects, due
to the interaction with the environment~\cite{HTK-12, MDPZ-12, CC-13,
  AKM-14}.  The competition between these two mechanisms may lead to
stationary states which are not related to a thermalization process,
i.e., whose properties are not describable in terms of thermal Gibbs
distributions.  In particular, novel phenomena may emerge close to a
quantum phase transition~\cite{Sachdev-book}, where the low-energy
properties of the system are particularly sensitive to variations of
the external conditions.

In this paper we investigate the dynamics of a many-body system in
proximity of a continuous quantum transition, in the presence of
dissipation arising from the interaction with the environment.  We
focus on a class of dissipative mechanisms whose dynamics can be
reliably described through a Lindblad master equation governing the
time evolution of the system's density matrix~\cite{BP-book, RH-book}.
This framework is of experimental interest, indeed the conditions for
its validity are typically realized in quantum optical and circuit-quantum
electrodynamics (c-QED) implementations~\cite{HTK-12, MDPZ-12,SBD-16}.
We argue that, in the presence of homogenous dissipators, the
competition between coherent and dissipative drivings develops dynamic
scaling laws involving some relevant parameters of the two mechanisms.
This occurs within a low-dissipation regime, where the decay rate of
the dissipation is comparable with the gap of the
Hamiltonian~\cite{NRV-19}.  General scaling behaviors are put forward,
which are expected to be developed along the time evolution described
by the Lindblad equation, especially in the large-time limit where
stationary states set in.  Analogously to the scaling laws of closed
systems at quantum transitions, the dynamic scaling behavior in the
presence of dissipation is expected to be universal, i.e., largely
independent of the microscopic details.

To verify the emerging scaling scenario, we found it convenient to
consider, as an example, the paradigmatic Kitaev quantum
wire~\cite{Kitaev-01}.  Namely, we study its dynamic behavior close to
the quantum transition, in the presence of dissipation due to local
incoherent pumping or decay. Our numerical results support the general
dynamic scaling theory, addressing the mutual interplay between
coherent dynamics and dissipation at a continuous quantum transition.

Some features arising from the competition of coherent and dissipative
dynamics close to quantum transitions were already recently analyzed
within a dynamic finite-size scaling framework, see
Ref.~\cite{NRV-19}, with specific emphasis on finite systems of linear
size $L$, and the dynamic behavior for relatively small times $t\sim
L^z$ (where $z>0$ is the dynamic critical exponent associated with the
quantum transition), thus not including the large-time stationary
regime.  In this paper we focus on some complementary regimes. Indeed
we consider infinite-size systems, for which we derive scaling laws
extending to the large-time limit of the evolution described by the
Lindblad equation, thus valid for the corresponding stationary states.
We shall emphasize that, at this stage, our scaling theory for quantum
dissipative systems should be considered as a conjecture arising
from phenomenological scaling arguments. Therefore, the scaling
behaviors devised for the two mentioned distinct situations
are not trivially related, and careful numerical checks are crucial
to validate our general framework.

We finally mention that dynamic scaling behaviors have been also put forward,
and numerically checked, for open critical systems when the
environment is constituted by a single qubit homogeneously coupled to
the whole many-body system~\cite{RV-19}.

The paper is organized as follows.  In Sec.~\ref{gensetup} we present
the general setup of our dynamic problem, recalling the scaling laws
developed by closed systems at quantum transitions, and the main
features of the dissipative mechanisms described by the Lindblad
equation.  Then, in Sec.~\ref{scalbeh} we put forward the scaling laws
describing the competition of coherent and dissipative drivings as
described by the Lindblad equation, extending the scaling laws of
closed systems at quantum transitions to incorporate the effects of
the dissipation.  In Sec.~\ref{Kitaevmo} we introduce the
one-dimensional Kitaev Fermi model in the presence of dissipation
arising from local pumping and decay, and define the observables that
we are going to consider, in order to characterize the time evolution
of the system, and in particular of the stationary states.
Section~\ref{numres} contains and discusses the outcomes of our
numerical results, which definitely support the dynamic scaling theory
in the presence of dissipation. Finally, in Sec.~\ref{conclu} we
summarize and draw our conclusions.

\section{Critical systems in the presence of dissipation}
\label{gensetup}

\subsection{Many-body systems at a quantum transition}
\label{masy}

We start by summarizing the general scaling features of a many-body
system at a continuous quantum transition.  Consider a generic
$d$-dimensional many-body system with Hamiltonian $\hat H$, close to a
zero-temperature transition driven by quantum
fluctuations~\cite{Sachdev-book, SGCS-97}.  A quantum transition is
generally characterized by few relevant perturbations, whose tuning
gives rise to quantum critical behaviors, characterized by a diverging
length scale, and universal power laws.

Let us assume that the system Hamiltonian has one relevant parameter
$\mu$, whose tuning toward the point $\mu_c$ develops a quantum
critical behavior.  The critical power laws are generally
characterized by the renormalization-group dimension $y_\mu$
associated with the relevant parameter $\mu$ and the dynamic exponent
$z$, so that the diverging length scale of the critical modes behaves
as
\begin{equation}
  \xi\sim \lambda \equiv |\bar{\mu}|^{-\nu}\,, \qquad
  \bar{\mu}\equiv \mu-\mu_c\,, \qquad \nu=y_\mu^{-1}\,,
  \label{xilandadef}
\end{equation}
and the suppression of the gap (that is, the difference of the two
lowest energy levels) as
\begin{equation}
  \Delta \sim \xi^{-z}\,.
  \label{zetadef}
\end{equation}
Moreover, the correlation function of generic local operators $\hat O(x)$, 
\begin{equation}
G(x;\bar{\mu}) \equiv \langle 0_\mu | \hat O(x)
\hat O(0) | 0_\mu\rangle
\label{gdefeq}
\end{equation}
where $|0_\mu\rangle$ is the ground state associated with the
parameter $\mu$, obeys an asymptotic scaling law of the
 form~\cite{SGCS-97,CPV-14}
\begin{equation}
  G(x; \, \bar{\mu}) \approx b^{-2y_o} {\cal G}(x/b,\bar{\mu}
  \,b^{y_\mu})\,,
  \label{gxbeq}
\end{equation}
where $b$ denotes an arbitrary positive number and $y_o$ is the
renormalization-group dimension of $\hat O(x)$. The above scaling
equation neglects further dependences on other irrelevant Hamiltonian
parameters, which are supposed to be suppressed in the large-$b$ limit.
Then, if we fix the arbitrary parameter $b$ by requiring 
\begin{equation}
  |\bar{\mu}| \, b^{y_\mu} =1
  \label{fixb}
\end{equation}
and introduce the variable $\lambda$ as in Eq.~\eqref{xilandadef}, we
obtain the asymptotic scaling behavior
\begin{equation}
  G(x; \, \bar{\mu}) \approx \lambda^{-2y_o} {\cal G}(x/\lambda)
  \label{gxscaeq}
\end{equation}
around $\mu_c$.  One may also consider equilibrium states at finite
temperature $T$, related to a Gibbs distribution of the quantum
states.  For sufficiently small temperatures, the dependence on $T$ is
taken into account~\cite{Sachdev-book} by adding a further dependence
on $T b^z$ in Eq.~(\ref{gxbeq}), turning into a dependence on
$T\lambda^z$ in the scaling equation (\ref{gxscaeq}), that is
\begin{equation}
  G(x; \, \bar{\mu},T) 
  \approx \lambda^{-2y_o} {\cal G}(x/\lambda,T\lambda^z)\,.
  \label{gxscaTeq}
\end{equation}
These asymptotic scaling behaviors are expected to be observed in the
limit $\lambda\to\infty$. Their approach is generally characterized by
power-law scaling corrections, which may come from different
sources~\cite{CPV-14}, such as irrelevant perturbations at the fixed
point describing the quantum transition, and analytic corrections in
the nonlinear scaling fields entering the scaling laws.

One may also study out-of-equilibrium evolutions close to a quantum
transition, for example arising from a sudden quench of the
Hamiltonian parameter $\mu$, at $t=0$ from an initial value $\mu_i$ to
$\mu\neq\mu_i$, starting from the ground state at $\mu_i$.  The
corresponding time dependence after the quench can be taken into
account within the scaling laws as well, by adding a further dependence on
$t\, b^{-z}$ in Eq.~(\ref{gxbeq}).  More precisely, the fixed-time
correlation functions are expected to behave as~\cite{PRV-18,NRV-18}
\begin{eqnarray}
  G(x,t; \, \bar{\mu}_i,\bar{\mu}) 
  &=& \langle \Psi(t) | \hat O(x) \hat O(0) |\Psi(t)\rangle\nonumber\\
  &\approx& \lambda^{-2y_o} 
  {\cal G}(x/\lambda,t\lambda^{-z},\bar{\mu}_i/\bar{\mu})\,,
  \label{gxscaoeq}
\end{eqnarray}
where $|\Psi(t)\rangle$ is the quantum state after a time $t$ from the
quench, and $\bar{\mu}_i \equiv \mu_i - \mu_c$.

The above scaling behaviors have been shown to develop for closed
states around the critical point $\mu_c$. In the following we study
the effects of the presence of dissipation in systems described by
critical Hamiltonians.

\subsection{Dissipative interactions}
\label{dissint}

Suppose that the many-body system is also subject to dissipative
interactions with the environment, so that the time dependence of its
density matrix $\rho$ is described by the Lindblad master
equation~\cite{BP-book}
\begin{equation}
  {\partial\rho\over \partial t} = -{i\over \hslash}[ \hat H,\rho]
  + u \,{\mathbb D}[\rho]\,,
  \label{lindblaseq}
\end{equation}
where the first term provides the coherent driving, while the second
term accounts for the coupling to the environment, characterized by a
global coupling constant $u>0$.

%%%%%%%%%%%%%%%%%%%%%%%%%%%%%%%%%%%%%%%%%%%%%%%%%%%%%%%%%%%%%%%%%%%%%%%
\begin{figure}[!b]
  \begin{tikzpicture}[scale=0.7]
    \draw [gray,thick] (1,0) --  (10,0);

%%%% Boundary conditions
    \draw [gray,thick,dashed] (0,0)--(1,0);
    \draw [gray,thick,dashed] (10,0)--(11,0);

    \foreach \i in {1,...,10}
{
        \filldraw [black] (\i-0.07,-0.07) rectangle ++(4pt,4pt); %circle (3pt);
        \draw[thick,<->,gray] (\i,-0.3)--(\i,-0.8);
        \draw [very thick, blue, rounded corners](\i-0.35,-1.8) rectangle (\i+0.35,-0.9);
        \filldraw[cyan!50!white, rounded corners] (\i-0.35,-1.8) rectangle (\i+0.35,-0.9);
        \node at (\i,-1.35) {$\mathcal{B}$};
}
  \end{tikzpicture}
  \caption{Sketch of a quantum system on a one-dimensional lattice
    (black squares), in which different sites may undergo a coherent
    and uniform nearest-neighbor coupling. Each site is supposed to be
    homogeneously and weakly coupled to an external and independent
    bath ${\mathcal B}$ (blue boxes), whose effect is to introduce
    local incoherent mechanisms.}
  \label{fig:sketch}
\end{figure}
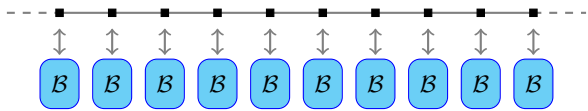
%%%%%%%%%%%%%%%%%%%%%%%%%%%%%%%%%%%%%%%%%%%%%%%%%%%%%%%%%%%%%%%%%%%%%%%

We restrict to homogeneous dissipation mechanisms, preserving
translational invariance, as sketched, for example, in
Fig.~\ref{fig:sketch}.  In the case of systems weakly coupled to
Markovian baths, the trace-preserving superoperator can be written as
a sum of local terms, such as~\cite{Lindblad-76,GKS-76}
\begin{eqnarray}
&&{\mathbb D}[\rho] = \sum_o {\mathbb D}_o[\rho]\,,\label{supop}\\
&&{\mathbb D}_o[\rho] = \hat L_o \rho \hat L_o^\dagger - \tfrac{1}{2}
  \big( \rho\, \hat L_o^\dagger \hat L_o + \hat L_o^\dagger \hat L_o
  \rho \big)\,,
  \label{dL}
\end{eqnarray}
where $\hat L_o$ is the Lindblad jump operator associated with the
local system-bath coupling scheme, and $o$ denotes an appropriate
spatial coordinate. Dissipation mechanisms described by {\em local}
Lindblad operators $L_o$ have been considered in various contexts (see,
e.g., Refs.~\cite{HTK-12, MDPZ-12, SBD-16, JBVMKFR-16, FSLKH-17}).
In several quantum optical devices, in particular in systems with
photon leakage or with some qubit relaxation or dephasing,
the conditions leading to Eqs.~\eqref{lindblaseq}-\eqref{dL} are typically
satisfied~\cite{HTK-12, MDPZ-12, SBD-16}, therefore this formalism constitutes
the standard choice for their theoretical investigation.
Other interesting implementations in this respect may be provided by hybrid
light-matter systems, where atoms are strongly coupled to cavities,
thus mediating photon-photon interactions, see e.g. Ref.~\cite{FSLKH-17}.

In the next section, we are going to address the formation of
large-time stationary states arising from the competition of the
coherent and dissipative drivings, as described by the Lindblad
equation~\eqref{lindblaseq}.  As we shall see below, while the
coupling with a bath generally drives the system to a noncritical
behavior, even when the Hamiltonian parameters are critical, it is
however possible to identify a peculiar low-dissipation regime where
the steady state may develop the above mentioned competition and
display a critical behavior.  To this purpose, we will put forward
general scaling behaviors for the stationary states in the
low-dissipation regime when the many-body Hamiltonian is within the
quantum critical regime, somehow extending the scaling laws reported
in Sec.~\ref{masy}.

\section{Scaling behavior of the stationary states}
\label{scalbeh}

Let us assume that, without loss of generality, the quantum many-body
system is initialized, at $t=0$, into the ground state of the
Hamiltonian $\hat{H}(\mu_i)$ for a given parameter $\mu_i$. The time
evolution for $t>0$ is dictated by the Lindblad equation
(\ref{lindblaseq}) with the Hamiltonian $\hat{H}(\mu)$, where $\mu$
may differ from $\mu_i$, thus realizing a sudden quench, and the
system-bath coupling strength is fixed by the coupling $u$.  The
dissipator ${\mathbb D}[\rho]$ may drive the system to a steady state,
which is generally noncritical, even when the Hamiltonian parameters
are critical.
In the case of dissipation leading to a unique stationary state, the
choice of the initial state (that is, the initial state of the time
evolution, which is eventually fixed by the parameter $\mu_i$) is not
relevant for the long-time properties of the system; nonetheless it
may determine the initial and intermediate time dependence.

We now argue that it is possible to identify a low-dissipation regime,
where the dissipation is sufficiently small to compete with the
coherent evolution driven by the (critical) Hamiltonian. This leads to
a late-time stationary state which can present a critical behavior,
depending also on the strength of the system-bath coupling.  The
effects of a small dissipation are taken into account by adding a
further dependence on a scaling variable associated with $u$ in the
out-of-equilibrium scaling laws, i.e.~$u \,b^{\zeta}$, where $\zeta$ is
a suitable exponent which ensures the substantial balance (thus
competition) with the critical coherent driving.  Since dissipation is
predicted to give rise to a relevant perturbation at the quantum
transition, we expect $\zeta>0$.  Thus, the peculiar low-dissipation
regime outlined above should be characterized by $u\sim
\lambda^{-\zeta}$.  

As already argued in Ref.~\cite{NRV-19}, the exponent $\zeta$ should
generally coincide with the dynamic exponent $z$.  This is expected by
noting that the parameter $u$ in Eq.~\eqref{lindblaseq} plays the role
of a decay rate, i.e., of an inverse relaxation time for the associated
dissipative process~\cite{BP-book}, and any relevant time scale $t_s$
at a quantum transition behaves as $t_s\sim
\Delta^{-1}$~\cite{PRV-18}.  In other words, to observe a nontrivial
competition between critical coherent dynamics and dissipation, one
should consider a sufficiently small coupling $u\sim \lambda^{-z}$, so
that its size is comparable with the energy difference $\Delta\sim
\lambda^{-z}$ of the lowest energy levels of the Hamiltonian.  An
analogous conjecture was put forward to describe the approach to
thermalization of some specific open systems close to a quantum
transition~\cite{YMZ-14}.  Here we extend it to more general
situations, even when the final stationary state is not thermal.

On the basis of such scaling arguments, generic fixed-time correlation
functions in the large-volume limit are thus expected to behave as
\begin{align}
  G(x,t; \, \bar{\mu}_i,\bar{\mu},u) &= {\rm Tr}\, \big[ \rho(t) \hat
    O(x) \hat O(0) \big] \nonumber\\ 
& \approx \lambda^{-2y_o} {\cal
    G}(x/\lambda,t\lambda^{-z},\bar{\mu}_i/\bar{\mu},u\lambda^z)\,,
  \label{gxscadiss}
\end{align}
where $\rho(t)$ is the time-dependent density matrix of the system.
We conjecture that this scaling ansatz describes the low-dissipation
regime of quenching protocols for many-body systems at quantum
transitions.  The main features of such dynamic critical regime are
expected to depend only on the universality class of the transition
and the general properties of the dissipative mechanism.

In order to derive the scaling laws for the asymptotic stationary
states, we need to consider the large-time limit of the scaling
equation (\ref{gxscadiss}). Hereafter we shall assume that the
asymptotic stationary state is unique, i.e., independent of the
initial conditions of the protocol that we consider, as is the case
for several classes of dissipators~\cite{Davies-70, Evans-77, SW-10,
  Nigro-19}.  Such stationary state should appear when $t\gg
\lambda^z$.  Therefore, we conjecture the scaling law
\begin{eqnarray}
  G_s(x; \, \bar{\mu},u) & \equiv &
  G(x,t\to\infty;\bar{\mu}_i,\bar{\mu},u) \nonumber\\
  &\approx& \lambda^{-2y_o} \, \Gamma(x/\lambda,u
  \lambda^{z})\,, \label{scagiti}
\end{eqnarray}
where in the definition of $G_s$ we assumed that the dependence
on the initial parameter $\bar{\mu}_i$ drops, under an assumption of
unicity of the stationary state.

It is also possible to define a correlation length $\xi_s$, from the
exponential large-distance decay of the correlation function
$G_s$. Indeed, since we expect that the large-distance behavior
$G_s\sim e^{-x/\xi_s}$, we may define
\begin{equation}
  \xi_s^{-1}(\bar{\mu},u) = -{\rm lim}_{x\to\infty} \, {{\rm ln} \,
    G_s(x;\bar{\mu},u)\over x} \,.
  \label{largedistdef}
\end{equation}
Using the scaling laws derived for $G_s$, we obtain
\begin{equation}
  \xi_s \approx \lambda \, {\cal L}_1(u \lambda^{z}) \,,
  \label{xiscai1}
\end{equation}
or alternatively
\begin{equation}
\xi_s \approx  u^{-1/z} \, {\cal  L}_2(u \lambda^{z})\,.
\label{xiscai2}
\end{equation}
Note that this scaling equation is formally analogous to that obtained
at the thermodynamic equilibrium, after replacing $u$ with the
temperature $T$, see Sec.~\ref{masy} and in particular
Eq.~(\ref{gxscaTeq}).  However, we stress that our arguments extend to
cases where the asymptotic stationary state does not coincide with a
Gibbs equilibrium state.  This will be indeed the case for the Kitaev
model subject to incoherent particle decay or pumping [see
  Eq.~\eqref{loppe} and the discussion below].

We finally stress that the strictly local (on-site) nature of the
Lindblad operators is not essential for the dynamic scaling behavior
put forward here.  The important point is that such interactions
only induce finite-size correlations, whose length scale becomes
negligible in the quantum critical limit, where critical quantum
correlations develop a diverging length scale in the system.

\section{The Kitaev quantum wire subject to dissipation}
\label{Kitaevmo}

To provide evidence of the scaling laws put forward in
Sec.~\ref{scalbeh}, we consider a Kitaev quantum wire defined by the
Hamiltonian~\cite{Kitaev-01}
\begin{equation}
  \hat H_{\rm K} = - t \sum_{j=1}^L \big( \hat c_j^\dagger \hat
  c_{j+1} + \delta \, \hat c_j^\dagger \hat c_{j+1}^\dagger+{\rm h.c.}
  \big) - \mu \sum_{j=1}^L \hat n_j \,,
  \label{kitaev2}
\end{equation}
where $\hat c_j$ is the fermionic annihilation operator on the $j$th
site of the chain, $\hat n_j\equiv \hat c_j^\dagger \hat c_j$ is the
density operator, and $\delta>0$.  We set $\hslash =1$, and $t=1$ as
the energy scale.

The Hamiltonian~\eqref{kitaev2} can be mapped into a spin-1/2 XY
chain, by means of a Jordan-Wigner transformation.  It undergoes a
continuous quantum transition at $\mu=\mu_c = -2$, independently of
$\delta$, between a disordered ($\mu<\mu_c$) and an ordered quantum
phase ($|\mu|<|\mu_c|$).  This transition belongs to the
two-dimensional Ising universality class~\cite{Sachdev-book},
characterized by the length-scale critical exponent $\nu=1$, related
to the renormalizaton-group dimension $y_\mu = 1/\nu=1$ of the
Hamiltonian parameter $\mu$ (more precisely of the difference
$\bar{\mu} \equiv \mu-\mu_c$).  The dynamic exponent associated with
the unitary quantum dynamics is $z=1$.  Moreover, the
renormalization-group dimension of the fermionic operators $\hat c_j$
and $\hat c^\dagger_j$ is $y_c = 1/2$, and that of the density
operator $\hat n_j$ is $y_n=1$~\cite{Sachdev-book}.

In the following we fix $\delta=1$, such that the corresponding spin model
is the quantum Ising chain
\begin{equation}
  \hat H_{\rm Is} = -\sum_j \hat
  \sigma^{(3)}_j \hat \sigma^{(3)}_{j+1} - g\, \sum_j \hat \sigma^{(1)}_j, 
  \label{isham}
\end{equation}
with $\hat \sigma^{(k)}_j$ being the Pauli matrices and $g=-\mu/2$.  In the
following we prefer to stick with the Kitaev quantum wire, because the
dissipation that we consider is more naturally defined for Fermi
lattice gases.

We focus on the dynamic behavior of the Fermi lattice
gas~\eqref{kitaev2} close to its quantum transition, in the presence
of homogeneous dissipation mechanisms following the Lindblad equation
(\ref{lindblaseq}). The dissipator ${\mathbb D}[\rho]$ is defined as a
sum of local (single-site) terms of the form
\begin{equation}
  {\mathbb D}_j[\rho] = \hat L_j \rho \hat L_j^\dagger - \tfrac{1}{2}
  \big( \rho\, \hat L_j^\dagger \hat L_j + \hat L_j^\dagger \hat L_j
  \rho \big)\,,
  \label{dLj}
\end{equation}
where $\hat L_j$ denotes the Lindblad jump operator associated with
the system-bath coupling scheme, and the index $j$ corresponds to a
lattice site [thus replacing the index $o$ in
  Eqs.~\eqref{supop},\eqref{dL}].  The onsite Lindblad operators $\hat
L_j$ describe the coupling of each site with an independent bath
${\mathcal B}$, Fig.~\ref{fig:sketch}. We consider dissipation
mechanisms associated with either particle losses (l)
or pumping (p)~\cite{HC-13, KMSFR-17}:
\begin{equation}
  \hat L_{{\rm l},j} = \hat c_j \,, \qquad \hat L_{{\rm p},j} = \hat
  c_j^\dagger \,.
  \label{loppe}
\end{equation}
The uniqueness of the eventual steady state has been proven for the
above decay and pumping operators~\cite{Davies-70, Evans-77, SW-10,
  Nigro-19}.  The choice of such dissipators turns out to be
particularly convenient for the numerically analysis, allowing us to
obtain results for quite large size of the Kitaev model, see below.

Our protocol starts from the ground state of $\hat H_K$ for a generic
$\bar{\mu}_i\equiv \mu_i-\mu_c$, sufficiently small to stay within the
critical regime.  To address the competition between coherent and
dissipative dynamics, we study the evolution after a quench of the
Hamiltonian parameter to $\bar{\mu}$, and a simultaneous sudden
turning on of the dissipation coupling $u$.  To characterize te
dynamic properties of the evolution described by the Lindblad
equation, and in particular the corresponding asymptotic large-time
behavior, we consider the fixed-time correlations
\begin{subequations}
\begin{eqnarray}
G_p(x,t) & \!\! = \!\! & {\rm Tr}[\rho(t)\,(\hat c_j^\dagger 
\hat c_{j+x}^\dagger +
    \hat c_{j+x} \hat c_{j})],\label{ptf}\\ 
G_c(x,t) & \!\! = \!\! & {\rm Tr}[\rho(t)\, (\hat c_j^\dagger \hat c_{j+x} 
+ \hat
    c_{j+x}^\dagger \hat c_{j})],\label{gtf}\\ 
G_n(x,t) & \!\! = \!\! & {\rm Tr}[\rho(t)\, \hat n_j \hat n_{j+x}] \! - \! 
{\rm Tr}[\rho(t)\, \hat n_j] \, {\rm Tr}[\rho(t)\, \hat n_{j+x}],
\qquad \; \label{gntf}
\end{eqnarray}
\end{subequations}
where we used the space translation invariance of the system.

According to the scaling arguments reported in Sec.~\ref{scalbeh} for
systems close to a continuous quantum transition, in particular for
$\lambda\equiv |\bar{\mu}|^{-1}\to\infty$, these correlation functions
are expected to develop the scaling laws given in
Eqs.~(\ref{gxscadiss}) and (\ref{scagiti}).  The approach to the
asymptotic behavior is foreseen to be controlled by power-law
corrections. Relying on the analysis of closed systems undergoing
quantum transition belonging to the two-dimensional Ising universality
class, see Ref.~\cite{CPV-14}, we generally expect $O(\lambda^{-1})$
corrections, generally arising from analytic corrections to the
nonlinear scaling fields [while scaling corrections arising from the
  leading irrelevant perturbation are more suppressed in the
  two-dimensional Ising universality class, as $O(\lambda^{-2})$].
The numerical results reported in the next section nicely support
these predictions.

\section{Numerical results}
\label{numres}

In this section we report numerical evidence of the scaling laws put
forward in Sec.~\ref{scalbeh} for quantum wires in the presence of
dissipation arising from incoherent particle losses and pumping.  The
choice of the dissipators (\ref{loppe}) allow us to numerically solve
the dynamic problem for systems with up to thousands of
sites~\cite{Prosen-08, Eisler-11}.  Indeed, the dynamics of the
dissipative fermionic Kitaev chain (Fig.~\ref{fig:sketch}) can be
written in terms of coupled linear differential equations, whose
number scales linearly with the size $L$.  We employ a fourth-order
Runge-Kutta method in order to integrate them.  Further details on
the computation of the time trajectories from the Lindblad
equation~\eqref{lindblaseq} are reported in Ref.~\cite{NRV-19}.

%%%%%%%%%%%%%%%%%%%%%%%%%%%%%%%%%%%%%%%%%%%%%%%%%%%%%%%%%%%%%%%%%%%%%%%
\begin{figure}[!t]
  \includegraphics[width=0.95\columnwidth]{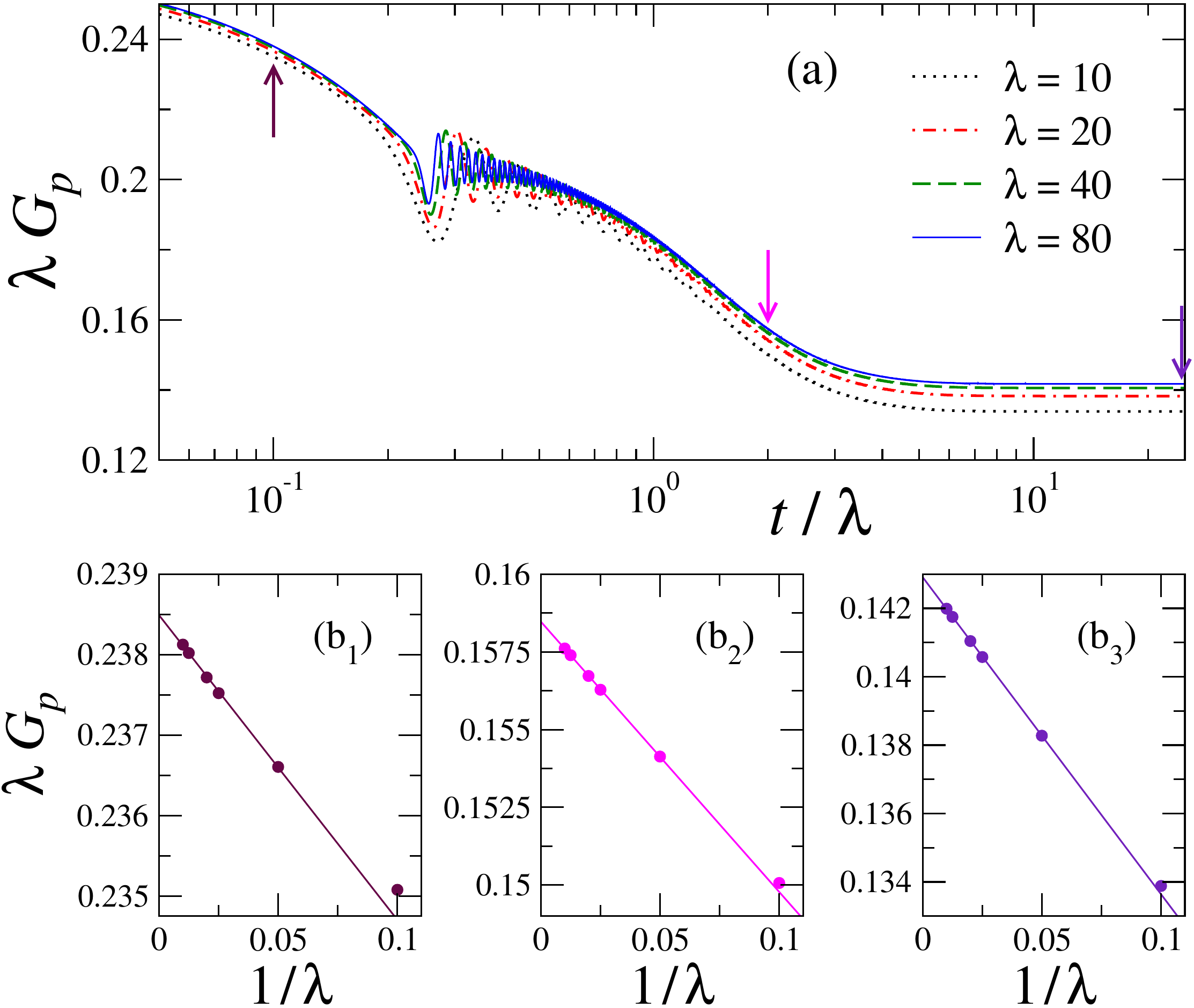}
  \caption{(a) The correlation function $G_p(x,t)$ in Eq.~\eqref{ptf},
    multiplied by $\lambda$, versus the rescaled time $t/\lambda$.
    Different curves are for various values of $\lambda$, as indicated
    in the legend.  Here we fix $x/\lambda=1$.  The dissipation is
    induced by incoherent particle losses and its rescaled strength is
    kept constant, such that $u \lambda=1$. We also fix $\bar \mu_i =
    \bar \mu$ (no Hamiltonian quenches) and take $\bar \mu < 0$ (i.e.,
    approaching the critical point from the left side: $\mu < \mu_c$),
    which vary for each curve as $|\bar \mu| = \lambda^{-1}$,
    cf.~Eq.~\eqref{xilandadef}.  The curves approach a scaling
    function with increasing $\lambda$, thus supporting the scaling
    equation~\eqref{gxscadiss}.  (b${}_1$)-(b${}_2$)-(b${}_3$) The
    convergence with $\lambda$ of the various curves in panel (a), for
    fixed values of the rescaled time, $t/\lambda = 0.1$, $2$, and
    $25$, respectively [arrows in panel (a)]. Straight lines are fits
    $\propto \lambda^{-1}$ of numerical data (circles), whose
    extrapolation to $\lambda \to +\infty$ produces the values 0.2385,
    0.1585, 0.1429, respectively.}
  \label{CorrP_X1_muNeg}
\end{figure}
%%%%%%%%%%%%%%%%%%%%%%%%%%%%%%%%%%%%%%%%%%%%%%%%%%%%%%%%%%%%%%%%%%%%%%%
 
%%%%%%%%%%%%%%%%%%%%%%%%%%%%%%%%%%%%%%%%%%%%%%%%%%%%%%%%%%%%%%%%%%%%%%%
\begin{figure}[!t]
  \includegraphics[width=0.95\columnwidth]{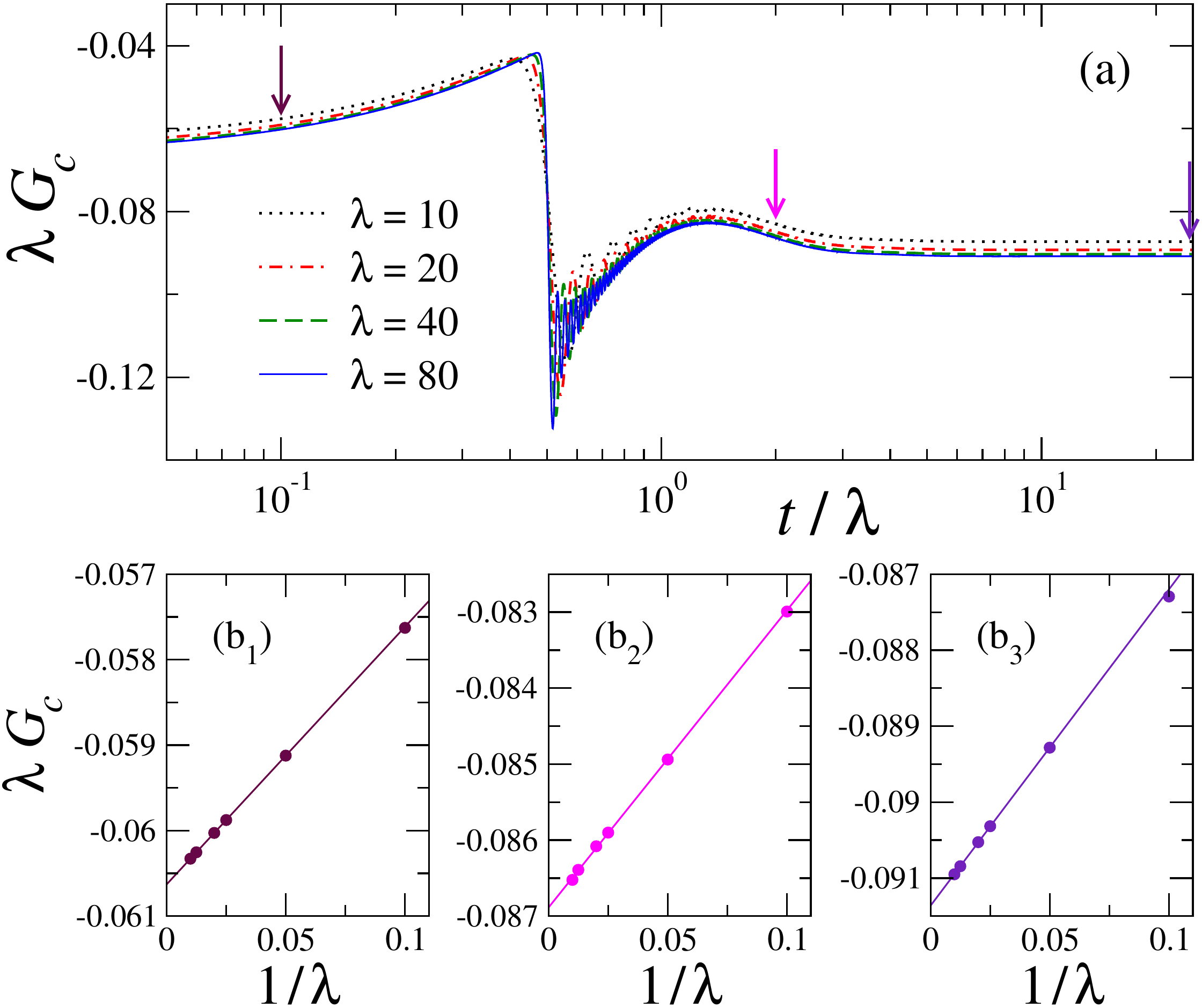}
  \caption{Same as in Fig.~\ref{CorrP_X1_muNeg}, but for the
    correlation function $G_c(x,t)$ in Eq.~\eqref{gtf}.  All the
    various parameters are the same as in
    Fig.~\ref{CorrP_X1_muNeg}, except for the value of $x/\lambda$,
    which here is taken fixed and equal to $2$.  The asymptotic values
    of the various curves in panel (a) for $\lambda \to +\infty$,
    extrapolated from a linear fit in $\lambda^{-1}$, are -0.0606
    [panel (b${}_1$), $t/\lambda=0.1$], -0.0869 [panel (b${}_2$),
      $t/\lambda=2$], -0.0914 [panel (b${}_3$), $t/\lambda=25$].}
  \label{CorrC_X2_muNeg}
\end{figure}
%%%%%%%%%%%%%%%%%%%%%%%%%%%%%%%%%%%%%%%%%%%%%%%%%%%%%%%%%%%%%%%%%%%%%%%

In our analysis we always consider antiperiodic boundary
conditions, i.e., such that $\hat c_{L+1} = - \hat c_1$, which turn
out to be technically convenient for the calculations.  However, since we always
consider sufficiently large chains to be effectively in the {\em
  thermodynamic} limit (more precisely $L\gg\lambda$, where
$\lambda=\bar\mu^{-1}$), the results that we present are not affected
by this particular choice.  The realization of the infinite-size limit is
easily verified by comparing the numerics at fixed $\lambda$ and $u$, with
increasing $L$.  All the data presented here should be considered in
the infinite-size limit with a great accuracy (finite-size effects are
invisible on the scales of all the figures shown below).

Let us start by presenting some numerical outcomes for the correlation
function $G_p(x,t)$, cf.~Eq.~\eqref{ptf}, in the presence of
dissipation arising from incoherent particle losses.  The curves shown
in Fig.~\ref{CorrP_X1_muNeg} have been obtained for the protocol with
$\bar \mu_i = \bar \mu$ (no Hamiltonian quenches) and $\bar \mu < 0$
(thus approaching the critical point from the left side: $\mu <
\mu_c$), keeping the rescaled dissipation strength constant, such that
$u \lambda^z =u\lambda=1$. The results in panel~(a) clearly support
the asymptotic dynamic scaling behavior put forward in
Eq.~(\ref{gxscadiss}), i.e., for $\mu_i=\mu$,
\begin{equation}
  G_p(x,t;\mu,u) \approx \lambda^{-1}\, 
  {\cal G}_p(x/\lambda, t/\lambda, u\lambda)\,.
  \label{scakigp}
\end{equation}
Indeed the data for $\lambda G_p(x,t)$ versus $t/\lambda$ at fixed
$x/\lambda=1$ appear to converge toward an asymptotic curve exhibiting
a nontrivial behavior, when increasing $\lambda$.  Notice also that
the long-time limit of such curve for $G_p$ is different
from zero, thus signaling the approach to a nontrivial stationary
state.  Moreover, as expected (see end of Sec.~\ref{Kitaevmo}), such
convergence is characterized by $O(\lambda^{-1})$ corrections, as is
visible from the bottom panels of Fig.~\ref{CorrP_X1_muNeg}, which
focus on three different values of $t/\lambda$: 0.1 [panel (b${}_1$)],
2 [panel (b${}_2$)], and 25 [panel (b${}_3$)].

We have extensively verified that the same scaling behavior of
Eq.~\eqref{scakigp} can be observed for any value of the ratio
$x/\lambda$, and of the rescaled dissipation strength $u\lambda$.
Furthermore an analogous dynamic scaling is developed by the
correlation function $G_c$ defined in Eq.~\eqref{gtf}, as explicitly
shown in Fig.~\ref{CorrC_X2_muNeg}.  For the sake of clarity in our
presentation, in that figure we adopted the same framework and
conventions of Fig.~\ref{CorrP_X1_muNeg}, and used the same set of
parameters, with the exception of $x/\lambda$, which here has been set
equal to $2$.  Also in that situation we notice a remarkable agreement
with the predicted $O(\lambda^{-1})$ scaling for finite-$\lambda$
corrections (see bottom panels of Fig.~\ref{CorrC_X2_muNeg}).

%%%%%%%%%%%%%%%%%%%%%%%%%%%%%%%%%%%%%%%%%%%%%%%%%%%%%%%%%%%%%%%%%%%%%%%
\begin{figure}[!t]
  \includegraphics[width=0.95\columnwidth]{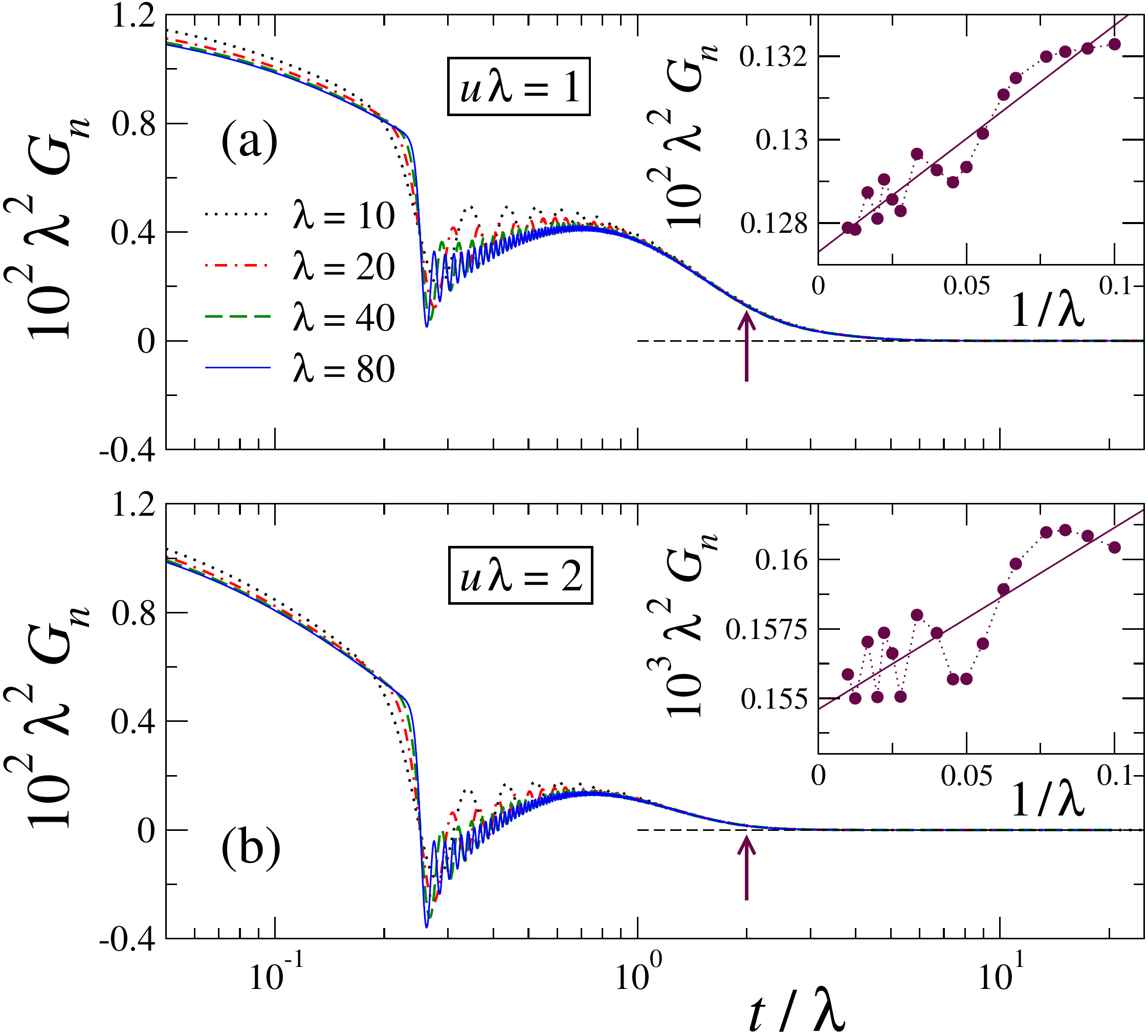}
  \caption{The correlation function $G_n(x,t)$ in Eq.~\eqref{gntf},
    multiplied by $\lambda^2$, versus the rescaled time $t/\lambda$.
    Panel (a) is with a dissipation strength $u \lambda = 1$, while
    panel (b) is with $u \lambda = 2$.
    Here we fix $x/\lambda=1$, $\bar \mu_i = \bar \mu$, and take
    $\bar \mu = - 1/ \lambda$. The
    horizontal dashed line indicates the zero value, which is reached
    asymptotically at long times, and is plotted as a guideline.  The
    two insets show the convergence with $\lambda$ of the curves at
    $t/\lambda = 2$ (arrow in the two main panels).  The rather
    complex behavior of the large-$\lambda$ convergence reflects the
    oscillatory behavior of the curves as a function of $t/\lambda$,
    which becomes more evident on the scale of the main frames for
    $0.3 \lesssim t/\lambda \lesssim 1$.  The dashed lines of the
    insets are drawn to guide the eyes (they suggest that the
    large-$\lambda$ limit approached by the data is $\approx 0.127$
    for $u \lambda = 1$ and $\approx 0.155$ for $u \lambda = 2$).}
  \label{CorrN_muNeg}
\end{figure}
%%%%%%%%%%%%%%%%%%%%%%%%%%%%%%%%%%%%%%%%%%%%%%%%%%%%%%%%%%%%%%%%%%%%%%%

As a further check of the dynamic scaling theory,
Fig.~\ref{CorrN_muNeg} displays results for the density-density
correlation function $G_n(x,t)$, cf.~Eq.~\eqref{gntf}, for two
different values of rescaled dissipation strength $u \lambda$.  Again,
they support the dynamic scaling behavior put forward in
Sec.~\ref{scalbeh}, see in particular Eq.~\eqref{gxscadiss}, keeping
into account that the renormalization-group dimension of the density
operator is now $y_n=1$.  Therefore, for $\bar{\mu}_i=\bar{\mu}$,
\begin{equation}
  G_n(x,t;\mu,u) 
  \approx \lambda^{-2}\, {\cal G}_n(x/\lambda, t/\lambda, u\lambda)\,,
  \label{scakign}
\end{equation}
i.e., the product $\lambda^2 G_n(x,t)$ is expected to approach a
nontrivial asymptotic large-$\lambda$ scaling behavior in terms of the
scaling variables $x/\lambda$, $t/\lambda$ and $u\lambda$, which
eventually vanishes for $t/\lambda \to \infty$.  The data in the two
insets, which focus on a fixed rescaled time $t/\lambda = 2$, reveal
that the $1/\lambda$ corrections to the scaling are, for this
observable, superimposed to fluctuations which witness the complex
oscillatory behavior of the scaling functions (compare the different
scales in the $y$-axis of the main panels and of the insets).

%%%%%%%%%%%%%%%%%%%%%%%%%%%%%%%%%%%%%%%%%%%%%%%%%%%%%%%%%%%%%%%%%%%%%%%
\begin{figure}[!t]
  \includegraphics[width=0.95\columnwidth]{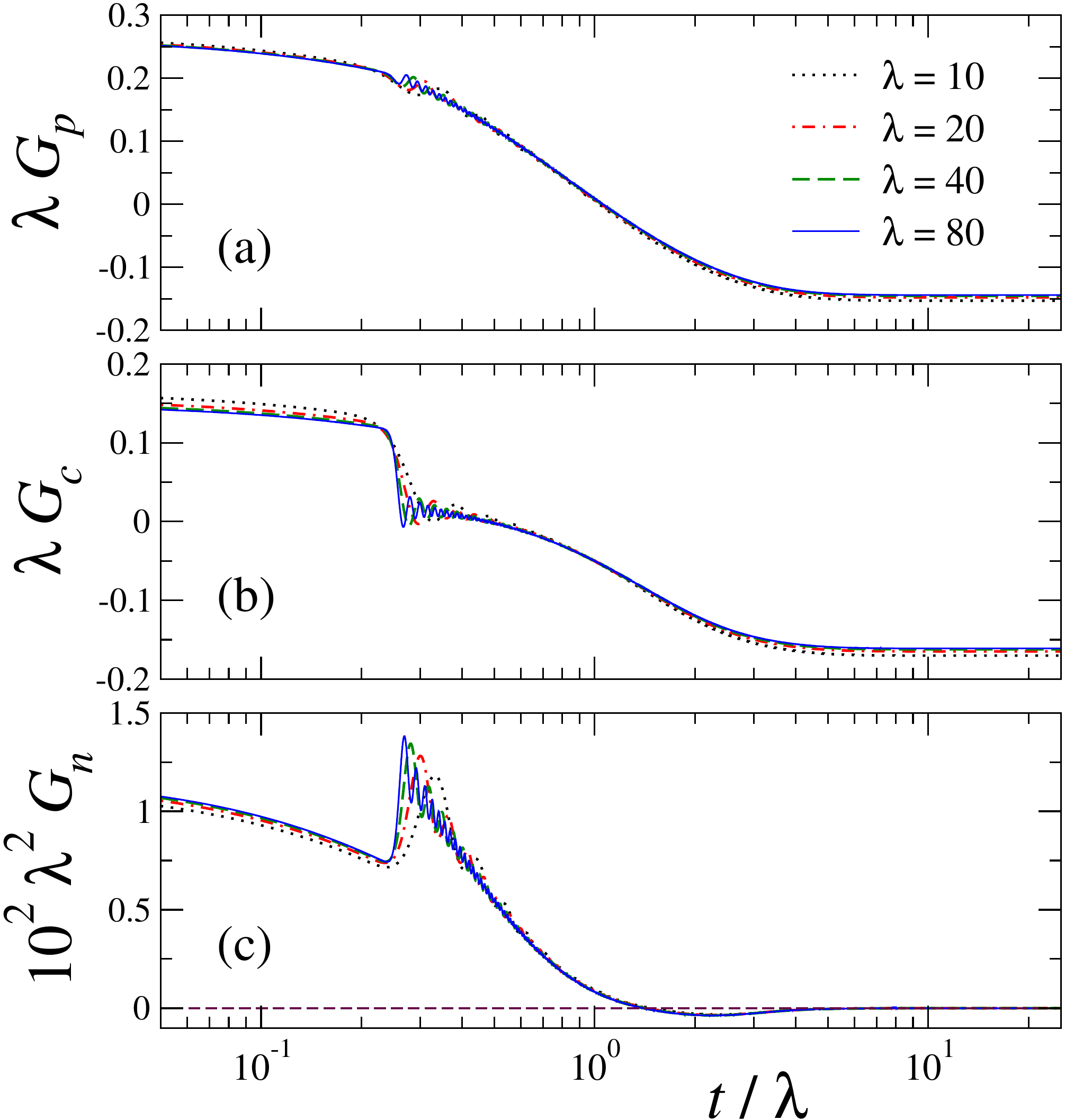}
  \caption{Results for $\lambda \, G_p(x,t)$ [panel (a)], $\lambda \,
    G_c(x,t)$ [panel (b)], and $\lambda^2 \, G_n(x,t)$ [panel (c)],
    versus the rescaled time $t/\lambda$.  The various system
    parameters are set as in Fig.~\ref{CorrN_muNeg}(a), with the
    exception of the Hamiltonian control parameter $\bar \mu_i = \bar
    \mu$, with $|\bar \mu| = \lambda^{-1}$, which here has been chosen
    to be positive (i.e., approaching the critical point from the
    right side: $\mu > \mu_c$).}
    \label{Corr_X1_muPos}
\end{figure}
%%%%%%%%%%%%%%%%%%%%%%%%%%%%%%%%%%%%%%%%%%%%%%%%%%%%%%%%%%%%%%%%%%%%%%%

Analogous scaling results can be obtained when approaching the
critical point from the right side: $\mu > \mu_c$, thus $\bar{\mu}>0$,
as shown by the curves reported in Fig.~\ref{Corr_X1_muPos} for the
three considered correlation functions.  Comparing the
results for the asymptotic stationary states with
those for $\mu < \mu_c$ at equal values of $|\bar{\mu}|$,
see, in particular,
  Fig.~\ref{CorrP_X1_muNeg}(a), Fig.~\ref{CorrC_X2_muNeg}(a), and
  Fig.~\ref{CorrN_muNeg}, we notice that
while the absolute values of the rescaled correlation functions
are the same when approaching the critical point from either side,
the respective signs for the correlation $G_c$ are exchanged.

%%%%%%%%%%%%%%%%%%%%%%%%%%%%%%%%%%%%%%%%%%%%%%%%%%%%%%%%%%%%%%%%%%%%%%%
\begin{figure}[!t]
  \includegraphics[width=0.95\columnwidth]{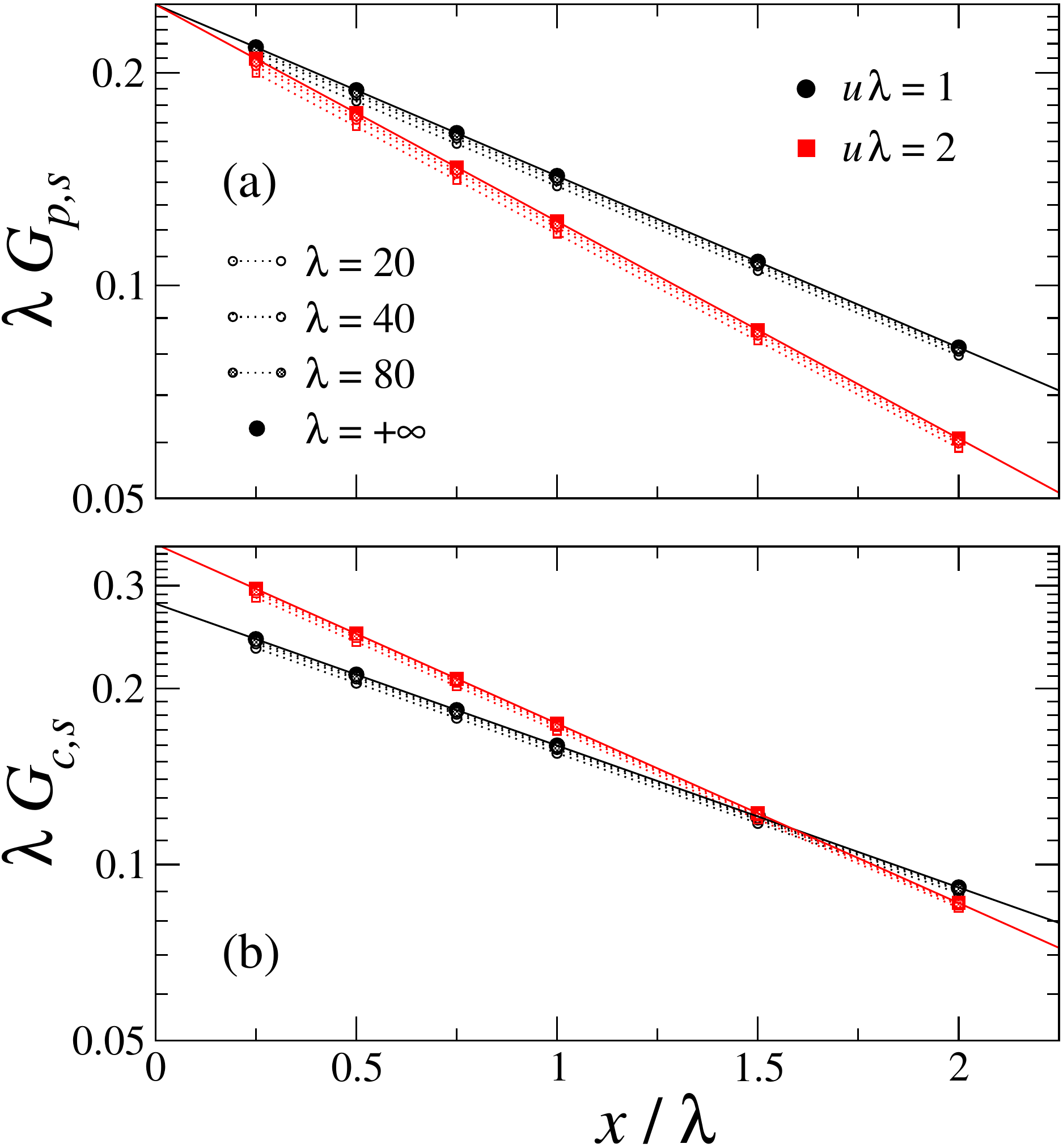}
  \caption{Results for $\lambda \, G_{p,s}(x)$ [panel (a)] and
    $\lambda \, G_{c,s}(x)$ [panel (b)] in the large-time limit
    (checked numerically with great accuracy), as a function of the
    rescaled variable $x/\lambda$.  Smaller and partially filled
    symbols stand for three specific values of $\lambda$, according to
    the legend, while fully filled symbols denote the values
    extrapolated for $\lambda \to +\infty$, that is, by extrapolating
    to zero the data at finite $\lambda^{-1}$ [see, e.g., type-(b)
      panels in Figs.~\ref{CorrP_X1_muNeg} and~\ref{CorrC_X2_muNeg}].
    Black data sets and circles are for a dissipation strength $u
    \lambda=1$, while red data sets and squares are for $u \lambda
    =2$.  Here we always choose $\bar \mu <0$.  Straight lines are
    exponential fits $G_s \sim e^{-(x/\lambda) / (\xi/\lambda)}$ of
    the extrapolated numerical data for $\lambda \to +\infty$.  For
    both kinds of correlation functions, the obtained decay rates are
    $\xi/\lambda = 1.789$ (for $u \lambda=1$) and $\xi/\lambda =
    1.414$ (for $u \lambda=2$), with a relative discrepancy smaller
    than $10^{-4}$, in support of the scaling law~\eqref{xiscai1}.}
  \label{Corr_asympt_PC}
\end{figure}
%%%%%%%%%%%%%%%%%%%%%%%%%%%%%%%%%%%%%%%%%%%%%%%%%%%%%%%%%%%%%%%%%%%%%%%

Let us now discuss more in detail the behavior of stationary states,
which are approached in the large-time limit.
Figure~\ref{Corr_asympt_PC} shows that the quantities $\lambda
G_{p,s}$ and $\lambda G_{c,s}$, obtained by taking the $t \to \infty$
limit of $G_p$ and $G_c$, approach an asymptotic large-$\lambda$
scaling form as a function of $x/\lambda$ and $u\lambda$
[cf.~Eq.~\eqref{scagiti}].  Indeed, they show that
\begin{equation}
  G_{p/c,s}(x; \mu_i,\mu,u) \approx
  \lambda^{-1}\, \Gamma_{p/c}(x/\lambda, u\lambda)\,,
  \label{gcpscaki}
\end{equation}
as put forward in Eq.~\eqref{scagiti}.
Moreover, they appear to decay exponentially for
sufficiently large distances, i.e.,
\begin{equation}
\lambda \, G_{p/c,s} \sim  \exp\left[-x/\xi_s\right] = 
    \exp\left[-\frac{x/\lambda}{\xi_s/\lambda}\right]\,,
\label{gsasy}
\end{equation}
where $\xi_s$ defines a correlation length which depends on the dissipation
strength, cf. Eq.~(\ref{largedistdef}). Therefore, Eq.~(\ref{gsasy}) also
implies the scaling behavior 
\begin{equation}
\xi_s\sim\lambda \equiv |\bar{\mu}|^{-\nu}\,,
\label{xisla}
\end{equation}
as predicted by the scaling equation~(\ref{xiscai1}).

\begin{figure}[!t]
  \includegraphics[width=0.95\columnwidth]{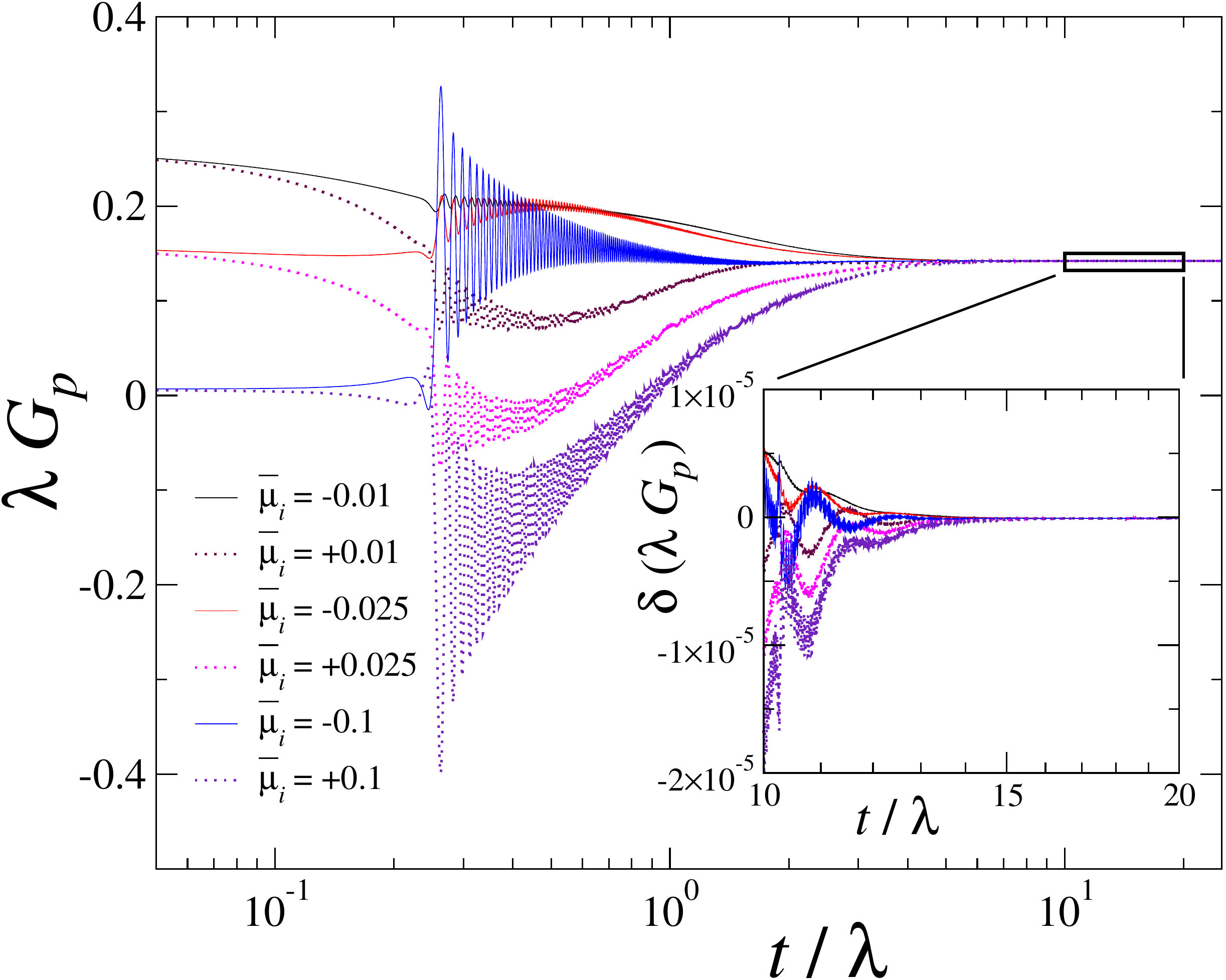}
  \caption{The correlation function $G_p(x,t)$ in Eq.~\eqref{ptf},
    multiplied by $\lambda$, versus the rescaled time $t/\lambda$.
    Here we fix $x/\lambda=1$, $u \lambda=1$, and $\bar \mu = -0.01$
    (corresponding to $\lambda = 100$).  The various curves correspond
    to different initial conditions, determined by the ground states
    of the Kitaev Hamiltonian with changing $\bar \mu_i$, being either
    negative (i.e., $\mu_i < \mu_c$) or positive ($\mu_i > \mu_c$)
    (see the legend).  The inset provides a magnification of the
    numerical data for $10 \leq t/\lambda \leq 20$ (box in the main
    frame), after subtracting the asymptotic value for $t/\lambda \to
    \infty$.}
  \label{Conv_initial}
\end{figure}

All the results presented so far have been obtained for $\mu_i = \mu$,
thus in the absence of any Hamiltonian quench.
Relaxing this assumption, we found that the stationary values
approached by the various correlators in the long-time limit
are independent of the initial condition of the protocol,
in particular of the value of $\mu_i\neq\mu$,
and thus obey the same asymptotic scaling behaviors discussed above. 
This is explicitly shown in Fig.~\ref{Conv_initial},
for the correlation function $G_p$.
This fact is consistent with the observation that, for our choice
of dissipators, the asymptotic stationary states are indeed unique.

%%%%%%%%%%%%%%%%%%%%%%%%%%%%%%%%%%%%%%%%%%%%%%%%%%%%%%%%%%%%%%%%%%%%%%%
%%%%%%%%%%%%%%%%%%%%%%%%%%%%%%%%%%%%%%%%%%%%%%%%%%%%%%%%%%%%%%%%%%%%%%%

We conclude our analysis by mentioning that completely analogous results
can be obtained in the case the dissipative mechanism is related
to a uniform local pumping, associated with the Lindblad operator
$\hat L_{{\rm p},j}$ in Eq.~(\ref{loppe}) (not shown).

\section{Summary and conclusion}
\label{conclu}

We have investigated the effects of dissipation on the dynamics of
many-body systems close to a continuous quantum phase transition,
arising from the interaction with the environment, as for example
sketched in Fig.~\ref{fig:sketch}. The latter is modeled
through a class of dissipative mechanisms that can be effectively
described by Lindblad equations for the density matrix
of the system~\cite{BP-book, RH-book}, with local and homogenous
Lindblad operators, such as those reported
in Eqs.~\eqref{lindblaseq}-\eqref{dL}.  This framework
is of experimental interest, indeed the conditions for its validity
are typically realized in quantum optical and c-QED
implementations~\cite{HTK-12, MDPZ-12, SBD-16}.

We have analyzed how homogenous dissipative mechanisms change the
scaling laws of closed systems at quantum transitions.  For this
purpose, we have considered a relatively simple dynamic protocol: the
quantum many-body system is initialized, at $t=0$, into the ground
state of the Hamiltonian $\hat H(\mu_i)$ for a given
parameter $\mu_i$; then for $t>0$ the system evolves according to the
Lindblad equation (\ref{lindblaseq}), where the coherent driving is
provided by the Hamiltonian $\hat{H}(\mu)$, and the dissipation
arising from the system-bath interaction is effectively described by the
dissipator ${\mathbb D}[\rho]$ with a fixed coupling strength $u$.

The large-time stationary state is usually noncritical, even when the
Hamiltonian parameters are critical.  However, we identified a
low-dissipation regime where the dissipation is sufficiently small to
compete with the coherent evolution driven by the critical
Hamiltonian, leading to stationary states which present critical
behaviors depending also on the strength of the dissipation coupling.
The above mentioned regime is generally realized when
the dissipation parameter $u$
scales as the gap $\Delta$ of the Hamiltonian of the many-body system,
i.e., $u\sim \Delta$.  Therefore it is a low-dissipation regime, in that
the gap gets suppressed at the quantum transition, as $\Delta\sim \xi^{-z}$,
where $\xi$ is the large length scale developed by the
critical correlation functions.  This reflects the fact that at a
quantum transition the perturbation arising from dissipation is always
relevant, such as the temperature at equilibrium~\cite{Sachdev-book,
  SGCS-97, CPV-14}.  This also means that, when $u\gg \Delta$, critical
fluctuations do not survive to the dissipation.

We argue that, under such low-dissipation conditions, open many-body
systems develop dynamic scaling laws, which apply to the time
evolution described by the Lindblad equation (\ref{lindblaseq}), and
in particular to the stationary states arising in the large-time limit
(see Sec.~\ref{scalbeh}).  Analogous to the scaling laws of closed
systems at quantum transitions, the dynamic scaling behavior in the
presence of dissipation is expected to be largely independent of the
microscopic details of both coherent and dissipative drivings, as for
the critical behavior of closed systems, which only depends on the
universality class of the quantum transition. Further investigation is
called for, to assess the actual extension of the universality of the
dynamic scaling functions with respect to the properties of the
dissipative mechanisms.

The dynamic scaling laws obtained in this paper apply to complementary
regimes with respect to those recently reported in Ref.~\cite{NRV-19},
where finite systems of linear size $L$ and dynamic behavior for
relatively small times $t\sim L^z$ were considered, thus not including
the large-time stationary regime.  On the other hand, here we focused
on infinite-size systems, for which we managed to derive scaling laws
extending to the large-time limit of the evolution described by the
Lindblad equation, thus valid for the corresponding stationary states.

The dynamic scaling scenario has been checked within fermion wires,
cf.  Eq.~\eqref{kitaev2}, in the presence of dissipation due to local
incoherent pumping and decay, which are described by the Lindblad
operators reported in Eq.~\eqref{loppe}.  Our numerical analysis
supports our general, phenomenological, dynamic scaling theory
addressing the competition between coherent dynamics and dissipation
at a continuous quantum transition.
Further checks of the dynamic scaling behaviors may turn
out to be interesting for other many-body systems at quantum
transitions, possibly belonging to different universality classes,
and/or dissipation mechanisms, including nonlocal ones~\cite{VWC-09,
  DRBZ-11, LPK-16}.

The arguments leading to this scenario are quite
general. We believe that analogous phenomena should develop in any
homogeneous $d$-dimensional many-body system at a continuous quantum
transition, whose Markovian interaction with the environment can be
described by local or extended dissipators within a Lindblad
equation~\eqref{lindblaseq}.  These arguments should also apply to
non-Markovian system-bath couplings~\cite{DA-17} (not described by
Lindblad equations), replacing $u$ with the parameter controlling the
decay rate.

We finally mention that some experimental breakthroughs were
recently achieved in the control of dissipative quantum many-body
dynamics, through different platforms, such as Rydberg atoms or
c-QED technology. Quantum critical behaviors in such
out-of-equilibrium context were reported in Refs.~\cite{CRWAW-13,
  TNDTT-17, FSLKH-17}.


\begin{thebibliography}{99}

\bibitem{HTK-12} A. A. Houck, H. E. T\"ureci, and J. Koch,
  On-chip quantum simulation with superconducting circuits,
  Nat. Phys. {\bf 8}, 292 (2012).

\bibitem{MDPZ-12}
  M. M\"uller, S. Diehl, G. Pupillo, and P. Zoller,
  Engineered open systems and quantum simulations with atoms and ions,
  Adv. At. Mol. Opt. Phys. {\bf 61}, 1 (2012).

\bibitem{CC-13}
  I. Carusotto and C. Ciuti, Quantum fluids of light,
  Rev. Mod. Phys. {\bf 85}, 299 (2013).
  
\bibitem{AKM-14}
  M. Aspelmeyer, T. J. Kippenberg, and F. Marquardt,
  Cavity optomechanics,  Rev. Mod. Phys. {\bf 86}, 1391 (2014).

\bibitem{Sachdev-book} S. Sachdev, {\em Quantum Phase Transitions},
  (Cambridge University, Cambridge, England, 1999).

\bibitem{BP-book} H.-P. Breuer and F. Petruccione, {\em The Theory of
  Open Quantum Systems} (Oxford University Press, New York, 2002).

\bibitem{RH-book} A. Rivas and S. F. Huelga, {\em Open Quantum
  System: An Introduction} (SpringerBriefs in Physics, Springer, 2012).
  
\bibitem{SBD-16} L. M. Sieberer, M. Buchhold, and S. Diehl, Keldysh
  field theory for driven open quantum systems, Rep. Prog. Phys. {\bf
    79}, 096001 (2016).

\bibitem{NRV-19}
  D. Nigro, D. Rossini, and E. Vicari,
  Competing coherent and dissipative dynamics close to quantum criticality,
  Phys. Rev. A {\bf 100}, 052108 (2019).

\bibitem{Kitaev-01} A. Yu. Kitaev, Unpaired Majorana fermions in
  quantum wires, Phys. Usp. {\bf 44}, 131 (2001).

\bibitem{RV-19} D. Rossini and E. Vicari, Scaling of decoherence and
  energy flow in interacting quantum spin systems, Phys. Rev. A
  {\bf 99}, 052113 (2019); E. Vicari, Decoherence dynamics of qubits
  coupled to systems at quantum transitions, Phys. Rev. A {\bf 98},
  052127 (2018).

\bibitem{SGCS-97} S. L. Sondhi, S. M. Girvin, J. P. Carini, and
  D. Shahar, Continuous quantum phase transitions,
  Rev. Mod. Phys. {\bf 69}, 315 (1997).

\bibitem{CPV-14} M. Campostrini, A. Pelissetto, and E. Vicari,
  Finite-size scaling at quantum transitions, Phys. Rev. B {\bf 89},
  094516 (2014).

\bibitem{PRV-18} A. Pelissetto, D. Rossini, and E. Vicari, Dynamic
  finite-size scaling after a quench at quantum transitions,
  Phys. Rev. E {\bf 97}, 052148 (2018). 

\bibitem{NRV-18} D. Nigro, D. Rossini, and E. Vicari, Scaling
  properties of work fluctuations after quenches near quantum
  transitions, J. Stat. Mech. (2019) 023104.

\bibitem{Lindblad-76}
  G. Lindblad, On the generators of quantum dynamical semigroups,
  Commun. Math. Phys. {\bf 48}, 119 (1976).

 \bibitem{GKS-76}
   V. Gorini, A. Kossakowski, and E. C. G. Sudarshan,
   Completely positive dynamical semigroups of N-level systems,
   J. Math. Phys. {\bf 17}, 821 (1976).

\bibitem{JBVMKFR-16}
  J. Jin, A. Biella, O. Viyuela, L. Mazza, J. Keeling, R. Fazio, and D. Rossini,
  Cluster mean-field approach to the steady-state phase diagram of dissipative
  spin systems,
  Phys. Rev. X {\bf 6}, 031011 (2016).

\bibitem{FSLKH-17} M. Fitzpatrick, N. M. Sundaresan, A. C. Y. Li,
  J. Koch, and A. A. Houck, Observation of a Dissipative Phase
  Transition in a One-Dimensional Circuit QED Lattice, Phys. Rev. X
  {\bf 7}, 011016 (2017).

\bibitem{YMZ-14} S. Yin, P. Mai, and F. Zhong, Nonequilibrium quantum
  criticality in open systems: The dissipation rate as an additional
  indispensable scaling variable, Phys. Rev. B {\bf 89}, 094108
  (2014); S. Yin, C.-Y. Lo, and P. Chen, Scaling behavior of quantum
  critical relaxation dynamics of a system in a heat bath,
  Phys. Rev. B {\bf 93}, 184301 (2016).

\bibitem{Davies-70} E. B. Davies, Quantum stochastic processes II,
  Commun. Math. Phys. {\bf 19}, 83 (1970); Quantum stochastic
  processes, Commun. Math. Phys. {\bf 15}, 277 (1969).

\bibitem{Evans-77} D. E. Evans, Irreducible Quantum Dynamical
  Semigroups, Commun. math. Phys. {\bf 54}, 293 (1977).

\bibitem{SW-10} S. G. Schirmer and X. Wang, Stabilizing open quantum
  systems by Markovian reservoir engineering, Phys. Rev. A {\bf 81},
  062306 (2010).

\bibitem{Nigro-19} D. Nigro, On the uniqueness of the steady-state
  solution of the Lindblad-Gorini-Kossakowski-Sudarshan equation,
  J. Stat. Mech. (2019) 043202.

\bibitem{HC-13} B. Horstmann, J. I. Cirac, and G. Giedke,
  Noise-driven dynamics and phase transitions in fermionic systems,
  Phys.Rev. A {\bf 87}, 012108 (2013).

\bibitem{KMSFR-17}
  M. Keck, S. Montangero, G. E. Santoro, R. Fazio, and D. Rossini,
  Dissipation in adiabatic quantum computers: lessons from an exactly
  solvable model, New. J. Phys. {\bf 19}, 113029 (2017).

\bibitem{Prosen-08} T. Prosen, Third quantization: a general method to
  solve master equations for quadratic open Fermi systems, New
  J. Phys. {\bf 10}, 043026 (2008).
  
\bibitem{Eisler-11} V. Eisler, Crossover between ballistic and
  diffusive transport: the quantum exclusion process,
  J. Stat. Mech. (2011) P06007.

\bibitem{VWC-09} F. Verstraete, M. Wolf, and J. I. Cirac, Quantum
  computation and quantum-state engineering driven by dissipation,
  Nat. Phys. {\bf 5}, 633 (2009).
  
\bibitem{DRBZ-11} S. Diehl, E. Rico, M. A. Baranov, and P. Zoller,
  Topology by dissipation in atomic quantum wires, Nat. Phys. {\bf 7},
  971 (2011).

\bibitem{LPK-16} A. C. Y. Li, F. Petruccione, and J. Koch, Resummation
  for nonequilibrium perturbation theory and application to open
  quantum lattices, Phys. Rev. X {\bf 6}, 021037 (2016).

\bibitem{DA-17}
  I. de Vega and D. Alonso, Dynamics of non-Markovian
  open quantum systems, Rev. Mod. Phys. {\bf 89}, 015001 (2017).

\bibitem{CRWAW-13}
  C. Carr, R. Ritter, C. G. Wade, C. S. Adams, and K. J. Weatherill,
  Nonequilibrium phase transition in a dilute Rydberg ensemble,
  Phys. Rev. Lett. {\bf 111}, 113901 (2013).

\bibitem{TNDTT-17}
  T. Tomita, S. Nakajima, I. Danshita, Y. Takasu, and Y. Takahashi,
  Observation of the Mott insulator to superfluid
  crossover of a driven-dissipative Bose-Hubbard system,
  Sci. Adv. {\bf 3}, e1701513 (2017).

\end{thebibliography}
\end{document}